\documentclass{revtex4}
\usepackage{amssymb}
\usepackage{amsfonts}
\usepackage{amsmath}

\input{tcilatex}

\begin{document}

\title{$n$-dimensional PDM non-linear oscillators: Linearizability and
Euler-Lagrange or Newtonian invariance}
\author{Omar Mustafa}
\email{omar.mustafa@emu.edu.tr}
\affiliation{Department of Physics, Eastern Mediterranean University, G. Magusa, north
Cyprus, Mersin 10 - Turkey,\\
Tel.: +90 392 6301378; fax: +90 3692 365 1604.}

\begin{abstract}
\textbf{Abstract:}\ We argue that, under multidimensional position-dependent
mass (PDM) settings, the Euler-Lagrange textbook invariance falls short and
turned out to be vividly incomplete and/or insecure for a set of
PDM-Lagrangians. We show that the transition from Euler-Lagrange component
presentation to Newtonian vector presentation is necessary and vital to
guarantee invariance. The totality of the Newtonian vector equations of
motion is shown to be more comprehensive and instructive than the
Euler-Lagrange component equations of motion (they do not run into conflict
with each other though). We have successfully used the \emph{Newtonian
invariance amendment}, along with some nonlocal space-time point
transformation recipe, to linearize Euler-Lagrange equations and extract
exact solutions for a set of $n$-dimensional nonlinear PDM-oscillators. They
are, Mathews-Lakshmanan type-I PDM-oscillators, power-law type-I
PDM-oscillators, the Mathews-Lakshmanan type-II PDM-oscillators, the
power-law type-II PDM-oscillators, and some nonlinear shifted
Mathews-Lakshmanan type-I PDM-oscillators.


\end{abstract}

\maketitle

\section{Introduction}

Classical and quantum mechanical particles endowed with position-dependent
mass (PDM) have initiated a substantial amount of research interest over the
last few decades \cite%
{1,2,3,4,5,6,7,8,9,10,11,12,13,14,15,16,17,18,19,20,21,22,23,24,25,26,27,28,29,30,31,32,33,34,35,36,37,38,39,40,41,42,43,44,45,46}%
. The position-dependent mass concept is, basically, a consequential
manifestation of either a position-dependent deformation in the standard
constant mass setting, or a position-dependent deformation in the
coordinates settings. A point mass moving within the curved
coordinates/space transforms, effectively, into\ a position-dependent mass
in Euclidean coordinates/space (c.f., e.g., \cite{3,5,20,45} and references
cited therein). Instantaneous Galilean invariance is used to derive the
PDM-Hamiltonian [46]. This would, in turn, offer a mathematically
challenging problem in both classical and quantum mechanics. In quantum
mechanics for example, the ordering ambiguity associated with the non-unique
representation of the PDM von Roos Hamiltonian \cite{1} possess a
significant amount of arguments as to what are the most proper parametric
ordering settings (e.g., \cite{2,3,4,5,6,7,8,9,10,11,12,13,14,15,16,17}). It
has been only recently that a proper definition for the position-dependent
mass momentum operator is introduced by Mustafa and Algadhi \cite{5},
resolving, hereby, the ordering ambiguity conflict. In classical mechanics,
nevertheless, exact solutions to multidimensional PDM Euler-Lagrange
equations are hard to find (e.g., \cite{18,19,20,21,30,33,38,44} and
references cited therein). One should, therefore seek some kind of nonlocal
space-time point transformations that guarantees Euler-Lagrange invariance
and facilitates exact solvability.

Based on the readily existing one-dimensional version \cite{38}, \ Mustafa
in \cite{44} has very recently embarked upon the $n$-dimensional extension
of the PDM Lagrangians via a nonlocal space-time point transformation and
sought invariance between the standard "constant" mass and PDM
Euler-Lagrange equations. Two $n$-dimensional PDM Lagrangian models were
used, 
\begin{equation}
L_{_{I}}\left( \vec{r},\vec{\upsilon};t\right) =\frac{1}{2}m_{\circ
}\sum\limits_{j=1}^{n}m_{_{j}}\left( x_{_{j}}\right) \,\dot{x}%
_{_{j}}^{2}-V_{_{I}}\left( \vec{r}\right) ;\text{ }\,\,\,j=1,2,\cdots ,n\in 
\mathbb{N}
,
\end{equation}%
and%
\begin{equation}
L_{_{II}}\left( \vec{r},\vec{\upsilon};t\right) =\frac{1}{2}m_{\circ
}m\left( \vec{r}\right) \sum\limits_{j=1}^{n}\dot{x}_{_{j}}^{2}-V_{_{II}}%
\left( \vec{r}\right) \,,\text{ }
\end{equation}%
where%
\begin{equation*}
\vec{r}=\sum\limits_{j=1}^{n}x_{_{j}}\,\hat{x}_{_{j}},\,\vec{\upsilon}=\frac{%
d\vec{r}}{dt}=\sum\limits_{j=1}^{n}\dot{x}_{_{j}}\,\hat{x}_{_{j}},\text{ }%
\dot{x}_{_{j}}=\frac{dx_{_{j}}}{dt}.
\end{equation*}%
Herewith, $m_{\circ }$ is the standard "constant" mass, $m_{_{j}}\left(
x_{_{j}}\right) $ in $L_{_{I}}\left( \vec{r},\vec{\upsilon};t\right) $\ is a
dimensionless scalar multiplier that deforms each coordinate $x_{_{j}}$
and/or velocity component $\dot{x}_{_{j}}$ in a specific functional form,
and $m\left( \vec{r}\right) $ in $L_{_{II}}\left( \vec{r},\vec{\upsilon}%
;t\right) $ represents a common dimensionless scalar multiplier that deforms
the coordinates $x_{_{j}}$'s and/or velocity components $\dot{x}_{_{j}}$'s.
Hence, similar consequential position-dependent deformations in the
potential force fields $V_{_{I}}\left( \vec{r}\right) $ and $V_{_{II}}\left( 
\vec{r}\right) $ are unavoidable in the process. Moreover, Mustafa \cite{44}
has considered a conventional constant-mass Lagrangian%
\begin{equation}
L\left( \vec{q},\overrightarrow{\tilde{q}};\tau \right) =\frac{1}{2}m_{\circ
}\sum\limits_{j=1}^{n}\tilde{q}_{_{j}}^{2}-V(\vec{q});\text{ \ }\tilde{q}%
_{_{j}}=\frac{dq_{_{j}}}{d\tau };\text{ }\,j=1,2,\cdots ,n,
\end{equation}%
where $\vec{q}=\left( q_{_{1}},q_{_{2}},\cdots ,q_{_{n}}\right) $ are some
generalized coordinates and $\tau $ is a re-scaled time. The idea is simply
a manifestation of Euler-Lagrange textbook invariance procedure. That is,
the Euler-Lagrange equations for $L_{_{I}}\left( \vec{r},\vec{\upsilon}%
;t\right) $ and $L_{_{II}}\left( \vec{r},\vec{\upsilon};t\right) $ should be
invariant with those for $L\left( \overrightarrow{q},\overrightarrow{\tilde{q%
}};\tau \right) $, if the used nonlocal space-time point transformation is
deemed useful.

However, it turned out that whilst the Euler-Lagrange equations for $%
L_{_{II}}\left( \vec{r},\vec{\upsilon};t\right) $ failed to satisfy
invariance conditions for $n\geq 2$, the Euler-Lagrange equations for $%
L_{_{I}}\left( \vec{r},\vec{\upsilon};t\right) $ proved to satisfy the
invariance conditions for $n\geq 1$. Such results would consequently render
the $n$-dimensional extension of the used nonlocal point transformation \cite%
{44} for the $L_{_{I}}\left( \vec{r},\vec{\upsilon};t\right) $ as a minor
and/or trivial progress. The said approach \cite{44} may\ very well copy and
paste the very recent work for the one-dimensional PDM-Lagrangians of
Mustafa \cite{38} (along with all examples discussed and reported therein)
for each degree of freedom $x_{i}$. For more details on this issue the
readers may refer to Mustafa \cite{44}. In the current methodical proposal,
therefore, we propose a remedy to this invariance problem in the form of the
what, hereinafter, should be called \emph{"Newtonian invariance amendment"}.
This is to be viewed as a comeback of the traditional textbook
Euler-Lagrange and Newtonian dynamical correspondence under PDM-settings.\
To the best of our knowledge, this has never been reported elsewhere in the
literature. The organization of our article is in order.

In section 2, we recycle, in short, the Euler-Lagrange equations invariance
for $L_{_{I}}\left( \vec{r},\vec{\upsilon};t\right) $ and $L_{_{II}}\left( 
\vec{r},\vec{\upsilon};t\right) $ with those for $L\left( \vec{q},%
\overrightarrow{\tilde{q}};\tau \right) .$ This would make the current
methodical proposal self-contained and vividly instructive. In the same
section, we introduce our \emph{Newtonian invariance amendment}. Hereby, we
show that while the textbook Euler-Lagrange invariance (for $n\geq 2$)
proved satisfactory only for $L_{_{I}}\left( \vec{r},\vec{\upsilon};t\right) 
$, the \emph{Newtonian invariance amendment} is found to be satisfactory for
both $L_{_{I}}\left( \vec{r},\vec{\upsilon};t\right) $ and $L_{_{II}}\left( 
\vec{r},\vec{\upsilon};t\right) $, $n\geq 1$ (when compared with the
conventional constant-mass Euler-Lagrange equations for $L\left( 
\overrightarrow{q},\overrightarrow{\tilde{q}};\tau \right) $, of course). We
may, therefore, label $L_{_{I}}\left( \vec{r},\vec{\upsilon};t\right) $ and $%
L_{_{II}}\left( \vec{r},\vec{\upsilon};t\right) $ as \emph{%
"target-Lagrangians"} and $L\left( \overrightarrow{q},\overrightarrow{\tilde{%
q}};\tau \right) $ as \emph{"reference Lagrangian"}. In fact, the total
vector presentation of Newtonian dynamical equation is nothings but the sum
of all Euler-Lagrange one-coordinate equations of motion, each of which is
multiplied by the corresponding unit vector (see (5), (6), (13) and (14)
below). Therefore, Newtonian invariance and Euler-Lagrange invariance do not
conflict each other. We devote section 3 to illustrate our methodical
proposal and use some conventional constant-mass $n$-dimensional oscillators
Lagrangian $L\left( \overrightarrow{q},\overrightarrow{\tilde{q}};\tau
\right) $ as a \emph{reference Lagrangian}. Where, three different
PDM-settings are successfully used within our nonlocal space-time point
transformation recipe. They are, $\vec{q}\left( \vec{r}\right) =\sqrt{%
m\left( \vec{r}\right) }\,\vec{r}$, $\vec{q}\left( \vec{r}\right) =\sqrt{%
m\left( \vec{r}\right) }\,\vec{\zeta}$ \ (where $\,\vec{\zeta}$ is a
constant vector), and $\vec{q}\left( \vec{r}\right) =\sqrt{m\left( \vec{r}%
\right) }\,\left( \vec{r}+\vec{\zeta}\right) $. Such vector transformational
recipes are mandatory substitutional settings for our \emph{Newtonian
invariance amendment}. We also show that the nonlinear PDM Euler-Lagrange
equations are linearizable through some nonlocal point transformations. In
the same section, moreover, some $n$-dimensional illustrative examples are
used. Amongst are, the Mathews-Lakshmanan type-I nonlinear PDM-oscillators
of (36), the power-law type-I nonlinear PDM-oscillators (40), the
Mathews-Lakshmanan type-II nonlinear PDM-oscillators (49), the power-law
type-II nonlinear PDM-oscillators (51), and some nonlinear shifted
Mathews-Lakshmanan type-I nonlinear PDM-oscillators (60). We conclude in
section 4.

\section{Newtonian invariance amendment to PDM Euler-Lagrange equations}

We start with recollecting/recycling some vital parts of the $n$-dimensional
extension of the PDM Lagrangians (via a nonlocal point transformation) work
by Mustafa \cite{44}. Therefore, we begin with the implementation of
Euler-Lagrange equations 
\begin{equation}
\frac{d}{dt}\left( \frac{\partial L}{\partial \dot{x}_{i}}\right) -\frac{%
\partial L}{\partial x_{i}}=0;\text{ }\,\,\,i=1,2,\cdots ,n\in 
\mathbb{N}
,
\end{equation}%
to obtain (with $m_{\circ }=1$ throughout) $n$ PDM Euler-Lagrange equations
( PDM EL-I) 
\begin{equation}
\ddot{x}_{_{i}}+\left( \frac{\dot{m}_{_{i}}\left( x_{i}\right) }{%
2m_{_{i}}\left( x_{i}\right) }\right) \,\dot{x}_{i}+\left( \frac{1}{%
m_{_{i}}\left( x_{i}\right) }\right) \,\partial _{x_{i}}V_{_{I}}\left( \vec{r%
}\right) =0;\text{ }\ddot{x}_{_{j}}=\frac{d^{2}x_{_{j}}}{dt^{2}},
\end{equation}%
for $L_{_{I}}\left( \vec{r},\vec{\upsilon};t\right) $, and PDM EL-II%
\begin{equation}
\ddot{x}_{_{i}}+\left( \frac{\dot{m}\left( \vec{r}\right) }{m\left( \vec{r}%
\right) }\right) \dot{x}_{i}-\frac{1}{2}\left( \frac{\partial
_{x_{i}}m\left( \vec{r}\right) }{m\left( \vec{r}\right) }\right)
\sum\limits_{j=1}^{n}\dot{x}_{_{j}}^{2}+\left( \frac{1}{m\left( \vec{r}%
\right) }\right) \,\partial _{x_{i}}V_{_{II}}\left( \vec{r}\right) =0,
\end{equation}%
for\ $L_{_{II}}\left( \vec{r},\vec{\upsilon};t\right) $, where $\partial
_{x_{i}}=\partial /\partial x_{i}$. Yet, the Euler-Lagrange equations for $%
L\left( \overrightarrow{q},\overrightarrow{\tilde{q}};\tau \right) $ ( EL-G)
yield%
\begin{equation}
\frac{d}{d\tau }\left( \frac{\partial L}{\partial \tilde{q}_{_{i}}}\right) -%
\frac{\partial L}{\partial q_{_{i}}}=0\Longleftrightarrow \frac{d}{d\tau }%
\tilde{q}_{_{i}}+\frac{\partial }{\partial q_{_{i}}}V(\vec{q})=0.
\end{equation}%
At this point, we shall seek some sort of feasible invariance for PDM EL-I
of (5) and for PDM EL-II of (6) with EL-G of (7). In so doing, Mustafa \cite%
{44} has suggested that we may extend/generalize the one-dimensional
nonlocal point transformation \cite{38} to fit into the $n$-dimensional
settings and re-scale both space and time through%
\begin{equation}
\text{ \ }d\tau =f\left( \vec{r}\right) dt,\,dq_{_{i}}=\delta _{ij}\sqrt{%
g\left( \vec{r}\right) }\,dx_{_{j}}\Longrightarrow \frac{\partial q_{_{i}}}{%
\partial x_{_{j}}}=\delta _{ij}\sqrt{g\left( \vec{r}\right) }.
\end{equation}%
This would necessarily mean that the unit vectors in the direction of $%
q_{_{i}}$ are obtained as 
\begin{equation}
\hat{q}_{_{i}}=\frac{\sum\limits_{k=1}^{n}\left( \frac{\partial x_{_{k}}}{%
\partial q_{_{i}}}\right) \hat{x}_{k}}{\sqrt{\sum\limits_{k=1}^{n}\left( 
\frac{\partial x_{_{k}}}{\partial q_{_{i}}}\right) ^{2}}}\Longrightarrow 
\hat{q}_{_{i}}=\,\hat{x}_{_{i}};\text{ }i=1,2,\cdots ,n.
\end{equation}%
Where the dimensionless functional structures of $f\left( \vec{r}\right) $
and $g\left( \vec{r}\right) $ shall be determined in the process below.

Under such settings, one obtains%
\begin{equation}
\text{\ }\tilde{q}_{_{j}}=\frac{\sqrt{g\left( \vec{r}\right) }}{f\left( \vec{%
r}\right) }\dot{x}_{_{j}}\,\Longrightarrow \frac{d}{d\tau }\text{\ }\tilde{q}%
_{_{j}}=\frac{\sqrt{g\left( \vec{r}\right) }}{f\left( \vec{r}\right) ^{2}}%
\left( \ddot{x}_{_{j}}+\dot{x}_{_{j}}\left[ \frac{\dot{g}\left( \vec{r}%
\right) }{2g\left( \vec{r}\right) }-\frac{\dot{f}\left( \vec{r}\right) }{%
f\left( \vec{r}\right) }\right] \right) ,
\end{equation}%
and\ hence EL-G of (7) would read%
\begin{equation}
\ddot{x}_{_{j}}+\left( \frac{\dot{g}\left( \vec{r}\right) }{2g\left( \vec{r}%
\right) }-\frac{\dot{f}\left( \vec{r}\right) }{f\left( \vec{r}\right) }%
\right) \dot{x}_{_{j}}+\left( \frac{f\left( \vec{r}\right) ^{2}}{g\left( 
\vec{r}\right) }\right) \,\partial _{x_{j}}V(\vec{q}\left( \vec{r}\right)
)=0;\,\,\,\,j=1,2,\cdots ,n.
\end{equation}%
Obviously, the invariance between EL-I of (5) and the resulting El-G of (11)
is feasible and is simply summarized by the relations%
\begin{equation}
\frac{d}{d\tau _{_{i}}}\left( \frac{\partial L}{\partial \tilde{q}_{_{i}}}%
\right) -\frac{\partial L}{\partial q_{_{i}}}=0\Longleftrightarrow \left\{ 
\begin{array}{c}
\partial q_{_{i}}/\partial x_{_{i}}=\sqrt{g_{_{i}}\left( x_{i}\right) }%
\smallskip \, \\ 
d\tau _{_{i}}\,/\medskip dt=f_{_{i}}\left( x_{i}\right) \\ 
\,g_{_{i}}\left( x_{i}\right) =m_{_{i}}\left( x_{i}\right) f_{_{i}}\left(
x_{i}\right) ^{2}\medskip \medskip \\ 
\text{ }V\left( \vec{r}\right) =V\left( \vec{q}\left( \vec{r}\right) \right)
\medskip \\ 
\smallskip \tilde{q}_{_{i}}\left( x_{i}\right) =\dot{q}_{_{i}}\left(
x_{i}\right) /f_{_{i}}\left( x_{i}\right) =\dot{x}_{_{i}}\sqrt{%
m_{_{i}}\left( x_{i}\right) }%
\end{array}%
\right\} \Longleftrightarrow \frac{d}{dt}\left( \frac{\partial L_{_{I}}}{%
\partial \dot{x}_{i}}\right) -\frac{\partial L_{_{I}}}{\partial x_{i}}=0.
\end{equation}%
However, it is clear that the dynamics of the $n$-dimensional PDM-system of $%
L_{_{I}}\left( \vec{r},\vec{\upsilon};t\right) $ in (1) fully decouples and
collapses into $n$ one-dimensional dynamical systems for each degree of
freedom $x_{i}$ (i.e., $n$ one-dimensional PDM EL-I equations of motion).
This would, in turn, render the $n$-dimensional extension proposal of
Mustafa \cite{44} as a minor and/or a trivial progress. For this approach,
of \cite{44}, may\ very well copy and paste our very recent work for the
one-dimensional PDM-Lagrangians in \cite{38} (along with all examples
discussed and reported therein) for each degree of freedom $x_{i}$. Whereas,
the comparison between El-G of (11) and PDM EL-II of (6) is only possible
for the one-dimensional problems (i.e., for $n=1$). In this case, (6)
collapses into (5) for $i=1=n$. Nevertheless. for the multidimensional case $%
n\geq 2$, the third term in (6) has no counterpart in (11). This would, in
effect, make the comparison incomplete/impossible and insecure. That is, for 
$n\geq 2$ the Euler-Lagrange equations (6) and (11) suggest that the
invariance is, apparently, still far beyond reach. Hereby, \emph{"Newtonian
invariance amendment"} comes into action.

Apriori, however, one should be reminded that a transition from
Euler-Lagrange into Newtonian dynamics is a simple textbook procedure and is
in order. Let us recollect the PDM EL-II of (6) and rephrase it to fit into
Newtonian vector dynamics. We do so by associating with each degree of
freedom a corresponding unit vector $\hat{x}_{_{i}}$ and sum over $%
i=1,2,\cdots ,n$ to get%
\begin{equation}
m\left( \vec{r}\right) \sum\limits_{i=1}^{n}\ddot{x}_{_{i}}\,\hat{x}_{_{i}}+%
\dot{m}\left( \vec{r}\right) \sum\limits_{i=1}^{n}\dot{x}_{_{i}}\,\hat{x}%
_{_{i}}-\frac{1}{2}\sum\limits_{i=1}^{n}\partial _{x_{i}}m\left( \vec{r}%
\right) \,\hat{x}_{_{i}}\left[ \sum\limits_{j=1}^{n}\dot{x}_{_{j}}^{2}\right]
+\,\,\sum\limits_{i=1}^{n}\hat{x}_{_{i}}\partial _{x_{i}}V_{_{II}}\left( 
\vec{r}\right) =0.
\end{equation}%
This would allow us to present the current equation in vector format
settings, with $\partial _{r}m\left( \vec{r}\right) =\partial m\left( \vec{r}%
\right) /\partial r$, as%
\begin{equation}
m\left( \vec{r}\right) \,\vec{a}+\partial _{r}m\left( \vec{r}\right) \left[ 
\frac{\,\vec{\upsilon}\,\left( \vec{r}\cdot \vec{\upsilon}\right) \,}{r}%
\right] -\frac{1}{2}\partial _{r}m\left( \vec{r}\right) \left[ \frac{\vec{r}%
\,\left( \vec{\upsilon}\cdot \vec{\upsilon}\right) }{r}\right] +\,\,\vec{%
\nabla}V_{_{II}}\left( \vec{r}\right) =0,\,
\end{equation}%
where, we have used%
\begin{equation}
\dot{m}\left( \vec{r}\right) =\sum\limits_{k=1}^{n}\partial _{x_{k}}m\left( 
\vec{r}\right) \,\dot{x}_{_{k}}=\frac{\partial _{r}m\left( \vec{r}\right) }{r%
}\sum\limits_{k=1}^{n}x_{_{k}}\dot{x}_{_{k}}=\frac{\partial _{r}m\left( \vec{%
r}\right) }{r}\,\left( \vec{r}\cdot \vec{\upsilon}\right) \,;\,r=\sqrt{%
\sum\limits_{j=1}^{n}x_{_{j}}^{2}}.
\end{equation}%
Obviously, equation (14) reduces to $m_{\circ }\,\vec{a}=-\,\,\vec{\nabla}%
V_{_{II}}\left( \vec{r}\right) $ for conventional constant mass settings
(i.e., for $m_{\circ }\neq 1$ and $m\left( \vec{r}\right) =1$). At this
point, nevertheless, let us consider two $n$-dimensional vectors $%
\overrightarrow{A}$ and $\overrightarrow{B}$ (i.e., $\overrightarrow{A},%
\overrightarrow{B}\in 
\mathbb{R}
^{n}$, where $%
\mathbb{R}
^{n}$ is the $n$-dimensional vector space). One may assert that, if $%
\overrightarrow{A}$ and $\overrightarrow{B}$ are parallel to each other
(i.e., co-directional $\overrightarrow{A}\parallel \overrightarrow{B}$) then 
$\overrightarrow{A}=\left\vert \overrightarrow{A}\right\vert \,\hat{A}%
=\left\vert \overrightarrow{A}\right\vert \hat{B}$ and $\overrightarrow{B}%
=\left\vert \overrightarrow{B}\right\vert \,\hat{B}=\left\vert 
\overrightarrow{B}\right\vert \,\hat{A}$. Consequently, the identity%
\begin{equation}
\left( \overrightarrow{A}\cdot \overrightarrow{B}\right) \overrightarrow{A}%
=\left( \overrightarrow{A}\cdot \overrightarrow{A}\right) \overrightarrow{B}%
\text{ };\text{ \ }\forall \overrightarrow{A}\parallel \overrightarrow{B}\in 
\mathbb{R}
^{n}
\end{equation}%
is intuitively most obvious and evident. This result would suggest that for
the case where $\vec{\upsilon}$ is parallel to $\vec{r}$ one obtains%
\begin{equation}
\,\vec{\upsilon}\,\left( \vec{\upsilon}\cdot \vec{r}\right) =\vec{r}\,\left( 
\vec{\upsilon}\cdot \vec{\upsilon}\right) \text{ ; }\,\vec{r}\,\parallel 
\vec{\upsilon},
\end{equation}%
to yield%
\begin{equation}
\,\vec{a}+\frac{1}{2}\frac{\partial _{r}m\left( \vec{r}\right) }{m\left( \,%
\vec{r}\right) }\left[ \frac{\vec{r}\,\left( \vec{\upsilon}\cdot \vec{%
\upsilon}\right) }{r}\right] +\frac{1}{m\left( \,\vec{r}\right) }\,\,%
\overrightarrow{\nabla }V_{_{II}}\left( \,\vec{r}\right) =0.\,
\end{equation}%
This would be acceptable for the case where no rotational effects are
involved (i.e., the case we are considering here), otherwise the Lagrangian
structure would include, in addition to the translational kinetic energy
term of (2), a rotational kinetic energy term (c.f., e.g., the
two-dimensional nonlinear oscillator kinetic energy term in \cite{29} and
equation (32) in \cite{20}) and a different treatment would be required,
therefore. Similarly, in a straightforward manner, one can show that EL-G of
(11) can be rewritten (in the Newtonian vector form) as%
\begin{equation}
\,\vec{a}+\left( \frac{\partial _{r}g\left( \vec{r}\right) }{2g\left( \vec{r}%
\right) }-\frac{\partial _{r}f\left( \vec{r}\right) }{f\left( \vec{r}\right) 
}\right) \left[ \frac{\vec{r}\,\left( \vec{\upsilon}\cdot \vec{\upsilon}%
\right) }{r}\right] +\left( \frac{f\left( \vec{r}\right) ^{2}}{\sqrt{g\left( 
\vec{r}\right) }}\right) \,\,\overrightarrow{\nabla }_{q}V(\vec{q}\left( 
\vec{r}\right) )=0\text{ ; }\,\,\overrightarrow{\nabla }_{q}=\sum%
\limits_{i=1}^{n}\hat{x}_{_{i}}\partial _{q_{i}}\text{ ; }\hat{q}_{_{i}}=\,%
\hat{x}_{_{i}}.
\end{equation}%
We got now consistency and exact correspondence between (18) and (19). That
is, the invariance between EL-G of (11) and PDM EL-II of (6) is now secured
and mandates that%
\begin{equation}
\frac{1}{2}\frac{\partial _{r}m\left( \vec{r}\right) }{m\left( \,\vec{r}%
\right) }=\frac{\partial _{r}g\left( \vec{r}\right) }{2g\left( \vec{r}%
\right) }-\frac{\partial _{r}f\left( \vec{r}\right) }{f\left( \vec{r}\right) 
}\Longleftrightarrow \frac{1}{m\left( \,\vec{r}\right) }=\frac{f\left( \,%
\vec{r}\right) ^{2}}{g\left( \,\vec{r}\right) }\Longleftrightarrow g\left( \,%
\vec{r}\right) =m\left( \,\vec{r}\right) f\left( \,\vec{r}\right) ^{2}.
\end{equation}%
Consequently, not only we have consistency between (18) and (19) but also we
have secured \emph{Newtonian invariance} between (6) and (7). We may,
therefore, safely rewrite (19) as%
\begin{equation}
\,\vec{a}+\frac{1}{2}\frac{\partial _{r}m\left( \vec{r}\right) }{m\left( \,%
\vec{r}\right) }\left[ \frac{\vec{r}\,\left( \vec{\upsilon}\cdot \vec{%
\upsilon}\right) }{r}\right] +\frac{1}{m\left( \,\vec{r}\right) }\,\,%
\overrightarrow{\nabla }_{q}V(\vec{q}\left( \vec{r}\right) )=0,
\end{equation}%
which immediately implies that $V_{_{II}}\left( \vec{r}\right) =V(\vec{q}%
\left( \vec{r}\right) )$. Wherein, we have used the relations%
\begin{equation}
\overrightarrow{\nabla }V_{II}\left( \vec{r}\right) =\sqrt{g\left( \vec{r}%
\right) }\,\overrightarrow{\nabla }_{q}V(\vec{q}\left( \vec{r}\right)
)\Longleftrightarrow \,\,\sum\limits_{i=1}^{n}\hat{x}_{_{i}}\partial
_{x_{i}}V_{_{II}}\left( \vec{r}\right) =\,\,\sum\limits_{i=1}^{n}\hat{x}%
_{_{i}}\partial _{x_{i}}V(\vec{q}\left( \vec{r}\right) )\Longleftrightarrow
V_{_{II}}\left( \vec{r}\right) =V(\vec{q}\left( \vec{r}\right) ).
\end{equation}%
This result is to be used to determine $q_{_{i}}\left( \vec{r}\right) $'s as
well as the form of $f(\vec{r})$ (consequently $g\left( \vec{r}\right) $)
for a given $m\left( \vec{r}\right) $. Moreover, in a straightforward
manner, the same procedure can be followed to show that the PDM EL-I is also
Newtonian invariant. Yet, we are now able to dismantle (21) into $n$
component equations%
\begin{equation}
\ddot{x}_{_{i}}+\left( \frac{\dot{m}\left( \vec{r}\right) }{2m\left( \vec{r}%
\right) }\right) \dot{x}_{i}+\left( \frac{1}{m\left( \vec{r}\right) }\right)
\,\partial _{x_{i}}V_{_{II}}\left( \vec{r}\right) =0;\text{ }i=1,2,\cdots ,n%
\text{,}
\end{equation}%
or%
\begin{equation}
\ddot{x}_{_{i}}+\frac{\partial _{r}m\left( \vec{r}\right) }{2\,r\,m\left( 
\vec{r}\right) }\left( \sum\limits_{j=1}^{n}\dot{x}_{_{j}}^{2}\right) x_{i}+%
\frac{1}{m\left( \vec{r}\right) }\,\partial _{x_{i}}V_{_{II}}\left( \vec{r}%
\right) =0;\text{ }i=1,2,\cdots ,n.
\end{equation}%
Where both forms hold true under our current settings. As such, our nonlocal
point transformation within our \emph{"Newtonian invariance amendment" } is
summarized by%
\begin{equation}
\frac{d}{d\tau }\left( \frac{\partial L}{\partial \tilde{q}_{_{i}}}\right) -%
\frac{\partial L}{\partial q_{_{i}}}=0\Longleftrightarrow \left\{ 
\begin{array}{c}
\hat{q}_{_{i}}=\,\hat{x}_{_{i}}\medskip \\ 
\partial q_{_{i}}/\partial x_{_{i}}=\sqrt{g\left( \vec{r}\right) }\,\text{\
\medskip } \\ 
d\tau \,=f\left( \vec{r}\right) \,dt\medskip \\ 
\,g\left( \vec{r}\right) =m\left( \vec{r}\right) f\left( \vec{r}\right)
^{2}\medskip \medskip \\ 
\text{ }V\left( \vec{r}\right) =V\left( \vec{q}\left( \vec{r}\right) \right)
\medskip \\ 
\tilde{q}_{_{i}}\left( \vec{r}\right) =\dot{q}_{_{i}}\left( \vec{r}\right)
/f\left( \vec{r}\right) =\dot{x}_{_{i}}\sqrt{m\left( \vec{r}\right) }\medskip%
\end{array}%
\right\} \Longleftrightarrow \frac{d}{dt}\left( \frac{\partial L_{_{II}}}{%
\partial \dot{x}_{i}}\right) -\frac{\partial L_{_{II}}}{\partial x_{i}}=0.
\end{equation}

We may now conclude that, whilst the conventional textbook Euler-Lagrange
invariance could only address the PDM EL-I settings (documented, in short,
above and in a sufficiently comprehensive details in \cite{44}) it could not
address the current PDM EL-II settings. Whereas, the what should be called,
hereinafter, \emph{"Newtonian invariance amendment" }\ works to perfection
for both PDM EL-I and PDM EL-II settings. One should also be reminded that 
\emph{"Newtonian invariance amendment"} is nothings but a manifestation of
the conventional constant mass Euler-Lagrange equations transition into the
Newtonian vector presentation of the equation of motion. However, the \emph{%
Newtonian invariance amendment} offered a vivid invariance to what is seemed
to be \emph{incomplete and/or insecure} invariance of the PDM Euler-Lagrange
equations. In the forthcoming illustrative examples, we clarify our
methodical proposal reported above.

\section{Nonlinear $n$-dimensional PDM-oscillators: Linearizability and
exact solvability}

Let us start with a constant mass $m_{\circ }$\ textbook harmonic oscillator
force field, in the generalized coordinates, 
\begin{equation}
V\left( \vec{q}\right) \medskip =\frac{1}{2}m_{\circ }\omega _{\circ
}^{2}\sum\limits_{j=1}^{n}q_{_{j}}^{2}=\frac{1}{2}m_{\circ }\omega _{\circ
}^{2}\left( \vec{q}\cdot \vec{q}\right) .
\end{equation}%
For which, one may use the EL-G equations of (7) to yield (with $m_{\circ
}=1 $) $n$ EL-G linear equations of motion%
\begin{equation}
\frac{d}{d\tau }\tilde{q}_{_{i}}+\omega _{\circ
}^{2}q_{_{i}}\,=0;\,i=1,2,\cdots ,n,
\end{equation}%
that admit exact solutions in the form of 
\begin{equation}
q_{_{i}}=B_{_{i}}\cos \left( \omega _{\circ }\tau +\varphi _{_{i}}\right) .
\end{equation}%
This is going to be our \emph{reference} case for the forthcoming \emph{%
target} PDM Lagrangians. In what follows, we shall seek forms of nonlocal
point transformations for $\vec{q}\left( \vec{r}\right) $ so that the
resulting nonlinear PDM Euler-Largrange equations admit linearizability into
(27). This would consequently suggest that one may use the exact solutions
(28) of (27) to extract exact solutions for the PDM- dynamical systems of
(23) or (24).

\subsection{Nonlinear $n$-dimensional PDM-oscillators: $\vec{q}\left( \vec{r}%
\right) =\protect\sqrt{m\left( \vec{r}\right) }\,\vec{r}$}

The substitutions of%
\begin{equation}
\vec{q}\left( \vec{r}\right) =\sqrt{m\left( \vec{r}\right) }\,\vec{r},
\end{equation}%
in (26), would imply the $n$-dimensional PDM-oscillators potential in the
form of%
\begin{equation}
V\left( \vec{q}\left( \vec{r}\right) \right) =V\left( \vec{r}\right) =\frac{1%
}{2}m\left( \vec{r}\right) \omega _{\circ
}^{2}\sum\limits_{j=1}^{n}x_{_{j}}^{2}.
\end{equation}%
Before we proceed any further, we need first to put the substitution (29) to
the test and see whether it satisfies our nonlocal point transformation
condition $\dot{q}_{_{i}}\left( \vec{r}\right) =\sqrt{m\left( \vec{r}\right) 
}f\left( \vec{r}\right) \dot{x}_{_{i}}$, of (25), or not. This is done in
order.%
\begin{equation}
\vec{q}\left( \vec{r}\right) =\sum\limits_{i=1}^{n}\hat{x}_{_{i}}q_{_{i}}=%
\sqrt{m\left( \vec{r}\right) }\,\sum\limits_{i=1}^{n}\hat{x}%
_{_{i}}x_{_{i}}\Longleftrightarrow \frac{d}{dt}\vec{q}\left( \vec{r}\right)
=\sum\limits_{i=1}^{n}\hat{x}_{_{i}}\dot{q}_{_{i}}=\sqrt{m\left( \vec{r}%
\right) }\left[ \vec{\upsilon}+\frac{\partial _{r}m\left( \vec{r}\right) }{%
2rm\left( r\right) }\vec{r}\,\left( \vec{r}\cdot \vec{\upsilon}\right) %
\right] .
\end{equation}%
We may now swap $\vec{r}$ and $\vec{\upsilon}$ in (17) and rewrite (31) as%
\begin{equation}
\sum\limits_{i=1}^{n}\hat{x}_{_{i}}\dot{q}_{_{i}}=\sqrt{m\left( \vec{r}%
\right) }\left[ \vec{\upsilon}+\frac{\partial _{r}m\left( \vec{r}\right) }{%
2rm\left( \vec{r}\right) }\,\vec{\upsilon}\,\left( \vec{r}\cdot \vec{r}%
\right) \right] =\sqrt{m\left( \vec{r}\right) }\left[ 1+\frac{r\,\partial
_{r}m\left( \vec{r}\right) }{2m\left( \vec{r}\right) }\,\right] \,\vec{%
\upsilon}\Longleftrightarrow \dot{q}_{_{i}}=\sqrt{m\left( \vec{r}\right) }%
\left[ 1+\frac{r\,\partial _{r}m\left( \vec{r}\right) }{2m\left( \vec{r}%
\right) }\,\right] \,\dot{x}_{_{i}}
\end{equation}%
Comparing this result with $\dot{q}_{_{i}}\left( \vec{r}\right) =\sqrt{%
m\left( \vec{r}\right) }f\left( \vec{r}\right) \dot{x}_{_{i}}$ of (25), we
obtain%
\begin{equation}
f\left( \vec{r}\right) =1+\frac{r\,\partial _{r}m\left( \vec{r}\right) }{%
2m\left( \vec{r}\right) }.
\end{equation}%
We, therefore, conclude that the substitution (29) satisfies our \emph{%
Newtonian invariance amendment} provided that $f\left( \vec{r}\right) $ is
given by (33). As such, equations (24) read%
\begin{equation}
\ddot{x}_{_{i}}+\frac{\partial _{r}m\left( \vec{r}\right) }{2\,r\,m\left( 
\vec{r}\right) }\left( \sum\limits_{j=1}^{n}\dot{x}_{_{j}}^{2}\right)
x_{i}+f\left( \vec{r}\right) \,\omega _{\circ }^{2}x_{_{i}}=0;\text{ }%
i=1,2,\cdots ,n,
\end{equation}%
which are our \emph{target} PDM Euler-Lagrange equations to be solved for
different PDM settings. Obviously, a reverse engineering of (34) into (27),
through the nonlocal transformation (25), would institute a linearization
process of (34) into (27). Under such nonlocal point transformation, we
consider two $n$-dimensional PDM-oscillator examples.

\subsubsection{Mathews-Lakshmanan type-I $n$-dimensional PDM-oscillators: $%
m\left( \vec{r}\right) =1/\left( 1\pm \protect\lambda r^{2}\right) $}

With a PDM function $m\left( \vec{r}\right) =1/\left( 1\pm \lambda
r^{2}\right) $, one can show that%
\begin{equation}
f\left( r\right) =1+\frac{r\,\partial _{r}m\left( \vec{r}\right) }{2m\left( 
\vec{r}\right) }=\frac{1}{1\pm \lambda r^{2}}.
\end{equation}%
Under such settings, the PDM Euler-Lagrange equations of (34) imply the
Mathews-Lakshmanan type-I $n$-dimensional PDM-oscillators equations of motion%
\begin{equation}
\ddot{x}_{_{i}}\mp \left( \frac{\lambda x_{_{i}}}{1\pm \lambda r^{2}}\right)
\,\sum\limits_{j=1}^{n}\dot{x}_{j}^{2}+\left( \frac{1}{1\pm \lambda r^{2}}%
\right) \,\omega _{\circ }^{2}x_{_{i}}=0;\text{ }i=1,2,\cdots ,n.
\end{equation}%
Which admit linearizability through the nonlocal transformation (25) and
inherit exact solutions of the form%
\begin{equation}
x_{_{i}}=B_{_{i}}\cos \left( \Omega t+\varphi \right) \text{ ; \ }\Omega
^{2}=\frac{\omega _{\circ }^{2}}{1\pm \lambda
\sum\limits_{i=1}^{n}B_{_{i}}^{2}}.
\end{equation}%
As such, the total energy is given by%
\begin{equation}
E=\frac{1}{2}\left( \frac{\omega _{\circ }^{2}}{1\pm \lambda
\sum\limits_{i=1}^{n}B_{_{i}}^{2}}\right) \sum\limits_{i=1}^{n}B_{_{i}}^{2}=%
\frac{1}{2}\Omega ^{2}\sum\limits_{i=1}^{n}B_{_{i}}^{2}
\end{equation}

\subsubsection{$n$-dimensional PDM power--law type-I oscillators: $m\left( 
\vec{r}\right) =k\,r^{2\protect\upsilon }$}

A power-law type PDM function $m\left( \vec{r}\right) =k\,r^{2\upsilon }$
would imply that%
\begin{equation}
f\left( r\right) =1+\frac{r\,\partial _{r}m\left( \vec{r}\right) }{2m\left( 
\vec{r}\right) }=1+\upsilon .\text{\ }
\end{equation}%
Hence, the PDM Euler-Lagrange equations of (34) yield the $n$-dimensional
PDM-oscillators equations of motion%
\begin{equation}
\ddot{x}_{_{i}}+\upsilon \,x_{_{i}}\,\left( \frac{\sum\limits_{j=1}^{n}\dot{x%
}_{_{j}}^{2}}{r^{2}}\right) +\left( 1+\upsilon \right) \,\omega _{\circ
}^{2}x_{_{i}}=0.
\end{equation}%
Such nonlinear equations of motion admit linearizability through the
nonlocal transformation (25) and inherit exact solutions of the form%
\begin{equation}
x_{_{i}}\,=C_{_{i}}\,\left[ \cos \left( \Omega t+\varphi _{_{i}}\right) %
\right] ^{1/\left( \upsilon +1\right) }\,;\text{ }\Omega =\left( 1+\upsilon
\right) \,\omega _{\circ },
\end{equation}%
where $\upsilon \neq -1,0$, otherwise trivial solutions are manifested.
Moreover, the total energy reads%
\begin{equation}
E=\frac{1}{2}\omega _{\circ }^{2}k\left(
\sum\limits_{i=1}^{n}C_{_{i}}^{2}\right) ^{\upsilon +1}=\frac{1}{2\left(
1+\upsilon \right) ^{2}}\Omega ^{2}k\left(
\sum\limits_{i=1}^{n}C_{_{i}}^{2}\right) ^{\upsilon +1}
\end{equation}

\subsection{Nonlinear $n$-dimensional PDM-oscillators: $\vec{q}\left( \vec{r}%
\right) =\protect\sqrt{m\left( \vec{r}\right) }\,\vec{\protect\zeta}$}

Let us now use the assumption that%
\begin{equation}
\vec{q}\left( \vec{r}\right) =\sqrt{m\left( \vec{r}\right) }\,\vec{\zeta}%
\text{ \ ; \ }\vec{\zeta}=\sum\limits_{j=1}^{n}\zeta _{_{j}}\hat{x}_{_{j}}%
\text{\ , }\zeta =\sqrt{\sum\limits_{j=1}^{n}\zeta _{_{j}}^{2}},
\end{equation}%
in (26), would imply the $n$-dimensional PDM-oscillators potential%
\begin{equation}
V\left( \vec{q}\left( \vec{r}\right) \right) =V\left( \vec{r}\right) =\frac{1%
}{2}m\left( \vec{r}\right) \,\omega _{\circ }^{2}\sum\limits_{j=1}^{n}\zeta
_{_{j}}^{2}=\frac{1}{2}m\left( \vec{r}\right) \,\omega _{\circ }^{2}\zeta
^{2}.
\end{equation}%
Where $\vec{\zeta}$ is a constant vector and is parallel to $\vec{r}\,$\ and 
$\vec{\upsilon}$ (i.e., $\vec{r}\,\parallel \vec{\upsilon}\parallel \vec{%
\zeta}$ ) and satisfies the vector identity in (17). Under such settings,
one would obtain%
\begin{equation}
\frac{d}{dt}\vec{q}\left( \vec{r}\right) =\sum\limits_{j=1}^{n}\hat{x}_{_{j}}%
\dot{q}_{_{j}}=\left( \frac{\dot{m}\left( \vec{r}\right) }{2\sqrt{m\left( 
\vec{r}\right) }}\right) \text{\ }\vec{\zeta}=\frac{\,\partial _{r}m\left( 
\vec{r}\right) }{2\,r\sqrt{m\left( \vec{r}\right) }}\left( \vec{r}\cdot \vec{%
\upsilon}\right) \,\vec{\zeta}=\frac{\,\partial _{r}m\left( \vec{r}\right) }{%
2\,r\sqrt{m\left( \vec{r}\right) }}\left( \vec{r}\cdot \vec{\zeta}\right) \,%
\vec{\upsilon}=\sqrt{m\left( \vec{r}\right) }\left[ \frac{\,\partial
_{r}m\left( \vec{r}\right) }{2\,m\left( \vec{r}\right) }\zeta \right] \,\vec{%
\upsilon},
\end{equation}%
which immediately implies that%
\begin{equation}
\sum\limits_{j=1}^{n}\hat{x}_{_{j}}\dot{q}_{_{j}}==\sqrt{m\left( \vec{r}%
\right) }\left[ \frac{\,\partial _{r}m\left( \vec{r}\right) }{2\,m\left( 
\vec{r}\right) }\zeta \right] \,\sum\limits_{j=1}^{n}\hat{x}_{_{j}}\dot{x}%
_{_{j}}\Longleftrightarrow \dot{q}_{_{j}}==\sqrt{m\left( \vec{r}\right) }%
\left[ \frac{\,\partial _{r}m\left( \vec{r}\right) }{2\,m\left( \vec{r}%
\right) }\,\zeta \right] \dot{x}_{_{j}}.
\end{equation}%
This result is consistent with $\dot{q}_{_{i}}\left( \vec{r}\right) =\sqrt{%
m\left( \vec{r}\right) }f\left( \vec{r}\right) \dot{x}_{_{i}}$ of (25)
provided that%
\begin{equation}
f\left( \vec{r}\right) =\frac{\,\partial _{r}m\left( \vec{r}\right) }{%
2\,m\left( \vec{r}\right) }\,\zeta .
\end{equation}%
Therefore, the PDM Euler-Lagrange equations of (24) read%
\begin{equation}
\ddot{x}_{_{i}}+\frac{\,\partial _{r}m\left( \vec{r}\right) }{2\,r\,m\left( 
\vec{r}\right) }\left( \sum\limits_{j=1}^{n}\dot{x}_{_{j}}^{2}\right) x_{i}+%
\frac{f\left( \vec{r}\right) }{r}\,\omega _{\circ }^{2}\,\zeta x_{_{i}}=0.
\end{equation}%
This result represents now our new \emph{target} PDM Euler-Lagrange equation
to be solved for different PDM settings. Yet again, a reverse engineering of
(48) into (27), through the nonlocal transformation (25), would institute a
linearization process of (48) into (27). Under such nonlocal point
transformation, we consider two $n$-dimensional PDM-oscillator examples.

\subsubsection{Mathews-Lakshmanan type-II $n$-dimensional PDM-oscillators: $%
m\left( \vec{r}\right) =1/\left( 1\pm \protect\lambda r^{2}\right) $}

With the substitution 
\begin{equation*}
m\left( \vec{r}\right) =\frac{1}{1\pm \lambda r^{2}}\Longleftrightarrow
f\left( \vec{r}\right) =\frac{\,\partial _{r}m\left( \vec{r}\right) }{%
2\,m\left( \vec{r}\right) }\,\zeta =\mp \frac{\lambda r}{1\pm \lambda r^{2}}%
\,\zeta ,
\end{equation*}%
the PDM Euler-Lagrange equations of (48) now read%
\begin{equation}
\ddot{x}_{_{i}}\mp \left( \frac{\lambda x_{_{i}}}{1\pm \lambda r^{2}}\right)
\,\sum\limits_{j=1}^{n}\dot{x}_{j}^{2}+\left( \frac{\mp \lambda \zeta ^{2}}{%
1\pm \lambda r^{2}}\right) \,\omega _{\circ }^{2}x_{_{i}}=0;\text{ }%
i=1,2,\cdots ,n.
\end{equation}%
Which is exactly the same as (36) provided that $\zeta ^{2}=\mp 1/\lambda $
(hence the notion \emph{"Mathews-Lakshmanan type-II }$n$\emph{-dimensional
PDM-oscillators"} is deemed appropriate). Such nonlinear equations of motion
admit linearizability through the nonlocal transformation (25) and inherit
the exact solutions of of (37) and (38).

\subsubsection{$n$-dimensional PDM power--law type-II oscillators: $m\left( 
\vec{r}\right) =\protect\lambda r^{2\protect\upsilon }$}

A power-law type PDM function $m\left( \vec{r}\right) =\lambda r^{2\upsilon
} $ would imply that%
\begin{equation}
f\left( \vec{r}\right) =\frac{\,\partial _{r}m\left( \vec{r}\right) }{%
2\,m\left( \vec{r}\right) }\,\zeta \Longleftrightarrow \frac{f\left( \vec{r}%
\right) }{r}=\frac{\upsilon \,\xi }{r^{2}},
\end{equation}%
and consequently the PDM Euler-Lagrange equations of (48) read%
\begin{equation}
\ddot{x}_{_{i}}+\frac{\upsilon }{r^{2}}\left( \sum\limits_{j=1}^{n}\dot{x}%
_{_{j}}^{2}\right) x_{i}+\frac{\upsilon \xi ^{2}}{r^{2}}\,\omega _{\circ
}^{2}\,x_{_{i}}=0;\text{ }i=1,2,\cdots ,n.
\end{equation}%
Such nonlinear equations of motion are linearizable into (27), through the
nonlocal transformation (25), and admit exact solutions in the form%
\begin{equation}
x_{_{i}}=B_{_{i}}\cos \left( \Omega t+\varphi \right) ;\text{ }\Omega ^{2}=%
\frac{\omega _{\circ }^{2}}{\lambda \left(
\sum\limits_{j=1}^{n}B_{_{j}}^{2}\right) },
\end{equation}%
provided that $\upsilon =-1$, and $\lambda =-1/\xi ^{2}$. Hence, the total
energy reads%
\begin{equation}
E=\frac{1}{2}\omega _{\circ }^{2}\xi ^{2}\left( \frac{\lambda }{%
\sum\limits_{j=1}^{n}B_{_{j}}^{2}}\right) =-\frac{1}{2}\Omega ^{2}\lambda =%
\frac{1}{2}\Omega ^{2}\xi ^{2}.
\end{equation}

\subsection{Nonlinear $n$-dimensional shifted PDM-oscillators: $\vec{q}%
\left( \vec{y}\right) =\protect\sqrt{m\left( y\right) }\,\vec{y},\,\vec{y}%
=\left( \vec{r}+\vec{\protect\zeta}\right) $}

A mixture of the two cases above, $\vec{q}\left( \vec{r}\right) =\sqrt{%
m\left( \vec{r}\right) }\,\vec{r}$ and $\vec{q}\left( \vec{r}\right) =\sqrt{%
m\left( \vec{r}\right) }\,\vec{\zeta}$ would imply that 
\begin{equation}
\vec{q}\left( \vec{y}\right) =\sqrt{m\left( \vec{y}\right) }\,\vec{y},\,\vec{%
y}=\left( \vec{r}+\vec{\zeta}\right) \Longleftrightarrow f\left( \vec{y}%
\right) =1+\frac{y\,\partial _{y}m\left( \vec{y}\right) }{2m\left( \vec{y}%
\right) }.
\end{equation}%
In this case, the oscillator potential of (26) yields an $n$-dimensional
shifted PDM oscillator potential 
\begin{equation}
V\left( \vec{q}\left( \vec{y}\right) \right) =V\left( \vec{y}\right) =\frac{1%
}{2}m\left( \vec{y}\right) \omega _{\circ
}^{2}\sum\limits_{j=1}^{n}y_{_{j}}^{2}=\frac{1}{2}m\left( \vec{y}\right)
\omega _{\circ }^{2}\sum\limits_{j=1}^{n}\left( x_{_{j}}+\xi _{_{j}}\right)
^{2}.
\end{equation}%
Under such shifted PDM settings, one may rewrite the transformation recipe
(25) as%
\begin{equation}
\frac{d}{d\tau }\left( \frac{\partial L}{\partial \tilde{q}_{_{i}}}\right) -%
\frac{\partial L}{\partial q_{_{i}}}=0\Longleftrightarrow \left\{ 
\begin{array}{c}
\hat{q}_{_{i}}=\,\hat{x}_{_{i}}=\hat{y}_{_{i}};\,\vec{y}=\sum%
\limits_{k=1}^{n}\left( x_{_{k}}+\xi _{_{k}}\right) \hat{x}_{_{k}}\medskip
\\ 
\partial q_{_{i}}/\partial y_{_{i}}=\sqrt{g\left( \vec{y}\right) }\text{\
\bigskip } \\ 
d\tau \,=f\left( \vec{y}\right) \,dt\medskip \\ 
\,g\left( \vec{y}\right) =m\left( \vec{y}\right) f\left( \vec{y}\right)
^{2}\medskip \medskip \\ 
\text{ }V\left( \vec{y}\right) =V\left( \vec{q}\left( \vec{y}\right) \right)
\medskip \\ 
\tilde{q}_{_{i}}\left( \vec{y}\right) =\dot{q}_{_{i}}\left( \vec{y}\right)
/f\left( \vec{y}\right) =\dot{y}_{_{i}}\sqrt{m\left( \vec{y}\right) }\medskip%
\end{array}%
\right\} \Longleftrightarrow \frac{d}{dt}\left( \frac{\partial L_{_{II}}}{%
\partial \dot{y}_{_{i}}}\right) -\frac{\partial L_{_{II}}}{\partial \dot{y}%
_{_{i}}}=0.
\end{equation}%
Which, consequently, suggests that the $n$-dimensional PDM Euler-Lagrange
equations are%
\begin{equation}
\ddot{x}_{_{i}}+\frac{\partial _{y}m\left( \vec{y}\right) }{2\,y\,m\left( 
\vec{y}\right) }\left( \sum\limits_{j=1}^{n}\dot{x}_{_{j}}^{2}\right) \left(
x_{_{i}}+\xi _{_{i}}\right) +f\left( \vec{y}\right) \,\omega _{\circ
}^{2}\left( x_{_{i}}+\xi _{_{i}}\right) =0,
\end{equation}%
where 
\begin{equation}
y=\sqrt{\sum\limits_{k=1}^{n}\left( x_{_{k}}+\xi _{_{k}}\right) ^{2}}.
\end{equation}%
It is obvious that for a PDM function of the form%
\begin{equation}
m\left( \vec{y}\right) =\frac{1}{1\pm \lambda y^{2}}=\frac{1}{1\pm \lambda
\sum\limits_{k=1}^{n}\left( x_{_{k}}+\xi _{_{k}}\right) ^{2}}%
\Longleftrightarrow f\left( \vec{y}\right) =m\left( \vec{y}\right) =\frac{1}{%
1\pm \lambda \sum\limits_{k=1}^{n}\left( x_{_{k}}+\xi _{_{k}}\right) ^{2}},
\end{equation}%
equations (57) would read the what may, very well, be called the \emph{%
shifted Mathews-Lakshmanan type-I }$n$\emph{-dimensional PDM-oscillators} 
\begin{equation}
\ddot{x}_{_{i}}\mp \left[ \frac{\lambda \left( x_{_{i}}+\xi _{_{i}}\right) }{%
1\pm \lambda \sum\limits_{k=1}^{n}\left( x_{_{k}}+\xi _{_{k}}\right) ^{2}}%
\right] \left( \sum\limits_{j=1}^{n}\dot{x}_{_{j}}^{2}\right) +\left( \frac{1%
}{1\pm \lambda \sum\limits_{k=1}^{n}\left( x_{_{k}}+\xi _{_{k}}\right) ^{2}}%
\right) \,\omega _{\circ }^{2}\left( x_{_{i}}+\xi _{_{i}}\right) =0;\text{ }%
i=1,2,\cdots ,n.
\end{equation}%
Such nonlinear equations of motion are linearizable into (27), through the
nonlocal transformation (56), and admit exact solutions in the form%
\begin{equation}
x_{_{i}}=A_{_{i}}\cos \left( \Omega t+\varphi \right) -\xi _{_{i}};\,\Omega
^{2}=\frac{\omega _{\circ }^{2}}{1\pm \lambda \sum\limits_{k=1}^{n}A_{k}^{2}}%
.
\end{equation}%
The total energy is then%
\begin{equation}
E=\frac{1}{2}\left( \frac{\omega _{\circ }^{2}}{1\pm \lambda
\sum\limits_{k=1}^{n}A_{k}^{2}}\right) \sum\limits_{i=1}^{n}A_{_{i}}^{2}=%
\frac{1}{2}\Omega ^{2}\sum\limits_{i=1}^{n}A_{_{i}}^{2},
\end{equation}%
which is in an obvious resemblance as that in (38).

\section{Concluding Remarks}

In this work, we have considered two-types of PDM-Lagrangians $%
L_{_{I}}\left( \vec{r},\vec{\upsilon};t\right) $ of (1) and $L_{_{II}}\left( 
\vec{r},\vec{\upsilon};t\right) $ of (2). They represent two different
PDM-Lagrangians structures. In $L_{_{I}}\left( \vec{r},\vec{\upsilon}%
;t\right) $, $m_{_{j}}\left( x_{_{j}}\right) $ is a dimensionless scalar
deformation in the coordinate $x_{_{j}}$ and/or velocity component $\dot{x}%
_{_{j}}$ in a specific functional form. Whereas, in $L_{_{II}}\left( \vec{r},%
\vec{\upsilon};t\right) $, $m\left( \vec{r}\right) $ is a common
dimensionless deformation for all coordinates $x_{_{j}}$ and/or velocity
components $\dot{x}_{_{j}}$. The feasibility of their textbook
Euler-Lagrange invariance with the conventional constant mass Lagrangians $%
L\left( \overrightarrow{q},\overrightarrow{\tilde{q}};\tau \right) $ of (3)
is studies via some $n$-dimensional nonlocal space-time point transformation
recipe of (8), (9), and (10). We have shown that, while the PDM
Euler-Lagrange equations for $L_{_{I}}\left( \vec{r},\vec{\upsilon};t\right) 
$ satisfy the invariance conditions with EL-G of $L\left( \overrightarrow{q},%
\overrightarrow{\tilde{q}};\tau \right) $ for $n\geq 1$, the PDM
Euler-Lagrange equations for $L_{_{II}}\left( \vec{r},\vec{\upsilon}%
;t\right) $ failed to do so for $n\geq 2$. This issue has stimulated and/or
inspired the current methodical proposal to introduce the new concept of 
{\footnotesize "}\emph{Newtonian invariance amendment"}.

As long as the conventional constant mass setting are in point, both
Euler-Lagrange invariance and \emph{Newtonian invariance }coincide with each
other. However, under the current PDM $n$-dimensional setting, it is deemed
necessary and vital that the transition from the Euler-Lagrange component
presentations, (6) and (11), to Newtonian vector presentations, (18) and
(19), should be carried out in order to secure invariance. It was obvious
that for $n\geq 2$ the invariance between Euler-Lagrange equations (6) and
(11) is still far beyond reach. Whereas, in the Newtonian presentations the
invariance between (18) and (19) is proved crystal clear. The totality of
the Newtonian vector presentation of the dynamical equation of motion is
shown to be more comprehensive/instructive than the Euler-Lagrange
components presentations. Hence, the notion {\footnotesize "}\emph{Newtonian
invariance amendment"} is rendered unavoidable for the current methodical
proposal. Yet, the \emph{Newtonian invariance amendment} has offered a vivid
invariance to what is seemed to be \emph{incomplete and/or insecure}
invariance of the PDM Euler-Lagrange equations. This is clarified and
documented in our analytical discussions in section 2. To the best of our
knowledge, this issue has never been reported elsewhere in the literature.

The $n$-dimensional linear Euler-Lagrange equations of motion (27) for the
oscillators Lagrangian%
\begin{equation}
L\left( \overrightarrow{q},\overrightarrow{\tilde{q}};\tau \right) =\frac{1}{%
2}m_{\circ }\sum\limits_{j=1}^{n}\tilde{q}_{_{j}}^{2}-\frac{1}{2}m_{\circ
}\omega _{\circ }^{2}\sum\limits_{j=1}^{n}q_{_{j}}^{2},
\end{equation}%
of (3) and (26), is used in section 3. Therein, we have used the
substitutions $\vec{q}\left( \vec{r}\right) =\sqrt{m\left( \vec{r}\right) }\,%
\vec{r}$, $\vec{q}\left( \vec{r}\right) =\sqrt{m\left( \vec{r}\right) }\,%
\vec{\zeta}$, and $\vec{q}\left( \vec{y}\right) =\sqrt{m\left( \vec{y}%
\right) }\,\vec{y}$\ \ \ (where $\vec{y}=\sum\limits_{k=1}^{n}\left(
x_{_{k}}+\xi _{_{k}}\right) \hat{x}_{_{k}}$) to find the corresponding $n$%
-dimensional nonlinear PDM-oscillators' Euler-Lagrange equations of motion
for the Lagrangians 
\begin{equation}
L_{_{II}}\left( \vec{r},\vec{\upsilon};t\right) =\frac{1}{2}m_{\circ
}m\left( \vec{r}\right) \sum\limits_{j=1}^{n}\dot{x}_{_{j}}^{2}-\frac{1}{2}%
m\left( \vec{r}\right) \omega _{\circ }^{2}\sum\limits_{j=1}^{n}x_{_{j}}^{2},
\end{equation}%
\begin{equation}
L_{_{II}}\left( \vec{r},\vec{\upsilon};t\right) =\frac{1}{2}m_{\circ
}m\left( \vec{r}\right) \sum\limits_{j=1}^{n}\dot{x}_{_{j}}^{2}-\frac{1}{2}%
m\left( \vec{r}\right) \omega _{\circ }^{2}\xi ^{2},
\end{equation}%
and%
\begin{equation}
L_{_{II}}\left( \vec{r},\vec{\upsilon};t\right) =\frac{1}{2}m_{\circ
}m\left( \vec{y}\right) \sum\limits_{j=1}^{n}\dot{x}_{_{j}}^{2}-\frac{1}{2}%
m\left( \vec{y}\right) \,\omega _{\circ }^{2}\sum\limits_{j=1}^{n}\left(
x_{_{j}}+\xi _{_{j}}\right) ^{2}.
\end{equation}%
Their Euler-Lagrange equations (34), (48), and (57), respectively, satisfy
the \emph{Newtonian invariance amendment}. Their linearizability into (27)
through some nonlocal transformations, (25) and (56), is discussed for
different PDM settings. We have used some illustrative examples that
include, the\emph{\ Mathews-Lakshmanan type-I PDM-oscillators} (36), the
power-law type-I PDM-oscillators (40), the \emph{Mathews-Lakshmanan type-II
PDM-oscillators} (49), the power-law type-II PDM-oscillators (51), and some
nonlinear \emph{shifted Mathews-Lakshmanan type-I PDM-oscillators} (60).
Hereby, using the exact solutions (28) of the standard $n$-dimensional
simple harmonic oscillator (27), we have successfully extracted exact
solutions for the PDM-dynamical systems in (36), (40), (49), (51), and (60).

Finally, we have not only introduced the new concept of the \emph{Newtonian
invariance amendment} as an alternative to Euler-Lagrange invariance, but
also we foresee that a new class of \emph{\ pseudo-superintegrable} and/or 
\emph{pseudo-superseparable} PDM-Lagrangians (and consequently
PDM-Hamiltonians) is implicitly introduced in the current methodical
proposal. The Lagrangian in (63), hence the corresponding Hamiltonian,
represent a class of\ \emph{superintegrable }Lagrangians/Hamiltonians in the
Liouville-Arnold sense of integrability (for more details on this issue the
reader may refer to the sample of references \cite{39,40,41,42,43,44,45} and
related references cited therein). That is, they introduce more constants of
motion (also called integrals of motion) than the degrees of freedom the
system is moving within. As long as the \ \emph{superintegrable }Lagrangian
(63), and its corresponding Hamiltonian, are transformable (through the
current nonlocal point transformation) into a set of
PDM-Lagrangians/Hamiltonians that do not even admit separability or
integrability, the descendent PDM-Lagrangians/Hamiltonians deserve to be
labeled as \emph{pseudo-superintegrable }and/or \emph{pseudo-superseparable}%
, therefore.

\end{document}